 %
 %
 %

 \ifx\mnmacrosloaded\undefined \input mn\fi
 %
 %
 %
 \input psfig
 \newif\ifAMStwofonts

 \ifCUPmtplainloaded \else
   \NewTextAlphabet{textbfit} {cmbxti10} {}
   \NewTextAlphabet{textbfss} {cmssbx10} {}
   \NewMathAlphabet{mathbfit} {cmbxti10} {} 
   \NewMathAlphabet{mathbfss} {cmssbx10} {} 
   \ifAMStwofonts
     \NewSymbolFont{upmath} {eurm10}
     \NewSymbolFont{AMSa} {msam10}
     \NewMathSymbol{\upi}     {0}{upmath}{19}
     \NewMathSymbol{\umu}     {0}{upmath}{16}
     \NewMathSymbol{\upartial}{0}{upmath}{40}
     \NewMathSymbol{\leqslant}{3}{AMSa}{36}
     \NewMathSymbol{\geqslant}{3}{AMSa}{3E}

      \let\le=\leqslant
      
   \else
     \def\umu{\mu}
     \def\upi{\pi}
     \def\upartial{\partial}
   \fi
 \fi



 \loadboldmathnames



 \letters          

 %
 %
 %
 %
 \def\PBvp #1 #2{ #1, #2}
 \def\PBa #1:#2 #3 #4 {#1,#2, {A\&A} \PBvp #3 #4}
 \def\PBapj #1:#2 #3 #4 {#1,#2, {ApJ} \PBvp #3 #4}
 \def\PBasupl #1:#2 #3 #4 {#1,#2, {A\&AS} \PBvp #3 #4}
 \def\PBapjsupl #1:#2 #3 #4 {#1,#2, {ApJS} \PBvp #3 #4}
 \def\PBpasp #1:#2 #3 #4 {#1,#2, { PASP} \PBvp #3 #4}
 \def\PBpaspc #1:#2 #3 #4 {#1,#2, { PASPC } \PBvp #3 #4}
 \def\PBmn #1:#2 #3 #4 {#1,#2, {MNRAS} \PBvp #3 #4}
 \def\PBmsait #1:#2 #3 #4 {#1,#2, {Mem. Soc. Astron. It.} \PBvp #3 #4}
 \def\PBnat #1:#2 #3 #4 {#1,#2, {Nat} \PBvp #3 #4}
 \def\PBaj #1:#2 #3 #4 {#1,#2, {AJ} \PBvp #3 #4}
 \def\PBjaa #1:#2 #3 #4 {#1,#2, {JA\& A} \PBvp #3 #4}
 \def\PBaspsc #1:#2 #3 #4 {#1,#2, {Ap\&SS} \PBvp #3 #4}
 \def\PBanrev #1:#2 #3 #4 {#1,#2, {ARA\&A} \PBvp #3 #4}
 \def\PBrevmex #1:#2 #3 #4 {#1,#2, {Rev. Mex. de Astron. y Astrof.} \PBvp #3 #4}
 \def\PBscie #1:#2 #3 #4 {#1,#2, {Sci} \PBvp #3 #4}
 \def\PBesomsg #1:#2 #3 #4 {#1,#2, {The Messenger} \PBvp #3 #4}
 \def\PBrmp #1:#2 #3 #4 {#1,#2, {Rev. Mod. Phys.} \PBvp #3 #4}
 \def\PBans #1:#2 #3 #4 {#1,#2, {Ann. Rev. of Nucl. Sci.} \PBvp #3 #4}
 \def\PBphrev #1:#2 #3 #4 {#1,#2, {Phys. Rev.} \PBvp #3 #4}
 \def\PBphreva #1:#2 #3 #4 {#1,#2, {Phys. Rev. A} \PBvp #3 #4}
 \def\PBphs #1:#2 #3 #4 {#1,#2, {Physica Scripta} \PBvp #3 #4}
 \def\PBjqsrt #1:#2 #3 #4 {#1,#2, {J. Quant. Spectrosc. Radiat.
        Transfer} \PBvp #3 #4}
 \def\PBcjp #1:#2 #3 #4 {#1,#2, {Can. J. Phys. } \PBvp #3 #4}
 \def\PBjphb #1:#2 #3 #4 {#1,#2, {J. Phys. B} \PBvp #3 #4}
 \def\PBapop #1:#2 #3 #4 {#1,#2, {Appl. Opt.} \PBvp #3 #4}
 \def\PBgca #1:#2 #3 #4 {#1,#2, {Geochim. Cosmochim. Acta}\PBvp #3 #4}
 %
 %

 %
 %

 \def\PBal{{ et al.~}}

 %
 %
 %
 %
 %
 \newcount\PBtn
 \def\PBcleartn{\global\PBtn=0}
 \def\PBtbl #1{\global\advance\PBtn by 1
 \begintable{\the\PBtn} 
 \caption{{\bf Table \the\PBtn .} #1}
 }
 \def\PBtbltwo #1{\global\advance\PBtn by 1
 \begintable*{\the\PBtn} 
 \caption{{\bf Table \the\PBtn .} #1}
 }

 \begintopmatter  

 \title{ Lithium  in very metal poor  thick disk stars}
 \author{
  P. Molaro$^1$, P. Bonifacio$^1$ and L. Pasquini$^2$}
 \affiliation{$^1$ Osservatorio Astronomico di Trieste, Via G.B. Tiepolo 11
 34131, Trieste -- Italy}
 \affiliation{$^2$ European Southern Observatory, K.  Schwarzschild Strasse 2
 D-85748 Garching bei Munchen -- Germany}

 \shorttitle{ Lithium in the thick disk}


 \PBcleartn

 \abstract { A search for lithium 
 is performed  on seven metal poor dwarfs with metallicities  ranging
  from [Fe/H]=-1.5 down to [Fe/H]=-3.0 but showing  
 disk--like kinematics. 
 These stars  belong to the   
 metal poor tail of the Galactic thick disk and they may be also
 the result of an accretion event 
 (Beers and Sommer-Larsen 1995). The Li 6707.8 \AA~ line
 is 
present in all the seven dwarfs.  The weighted average of the Li abundance
 for the stars is  A(Li)=2.20 $\pm 0.06$ and  is  consistent 
within the errors
 with the {\it plateau} Li abundance  of A(Li)=2.24$\pm$ 0.012 found in 
genuine   halo stars in the same range of
 metallicities (Bonifacio \& Molaro 1997). One of the stars,
CS 22182-24, shows 
somewhat lower Li abundance 
(A(Li)=1.6$\pm$0.40) and is a candidate to being a Li-poor star.
   Whether  this group of stars belongs to   
 the oldest stars in the disk or  to 
the old population of an external galaxy
 accreted by the Milky Way, the present observations provide  support
 to  the  universality of a   pre-Galactic  Li abundance 
  as  is observed in the Galactic halo stars. 
 }


 \maketitle  

 \section{Introduction}

 Spite \& Spite (1982) discovered  the presence of Li  in the
 warm halo  dwarfs at a constant value (the Spite {\it plateau})
 and suggested that it  is the
  primordial Li, i.e. Li synthesized in the Big Bang.  
 The knowledge of primordial 
Li   is  of far reaching  cosmological  relevance
  since 
 it  is a function of $\eta\equiv n_b/n_\gamma$, the
 baryon to photon ratio, which is related to total baryonic density 
by $\Omega_{B} h^2=0.0066 \eta_{10}$.  
 The Spite \& Spite   findings have  
 been  confirmed by a number of investigations and  
  Li is now measured in somewhat  one hundred  
of halo dwarfs (see Ryan et al 1996 and references 
therein) including  stars with
  metallicities down to [Fe/H]$\approx -4.0 $ and
 few turn-off globular 
 cluster stars (Molaro \& Pasquini 1994; Deliyannis et al 1995;
Pasquini 
\& Molaro 1997). 
 Only very few stars  have been found with a Li abundance much less
 than the 
 {\it plateau} value. 
The substantial flatness of the {\it plateau}  and the
remarkable  absence of
  any  scatter in the Li abundance   have been recently confirmed by means
 of an accurate  analysis of a  
 subsample of stars   with accurate 
 effective temperatures (Bonifacio \& Molaro 1997). 
 These properties together with 
 the detection of the more fragile isotope $\rm ^6Li$ in 
 HD 84937 (Smith et al 1993 Hobbs \& Thorburn 1997) are   the main arguments 
 to reject  significant modification  
 of the pristine Li abundance due to stellar
 and/or Galactic processes.

 So far the Li measurements  
generically refer to halo stars regardless of their
 belonging to the {\it inner} or {\it 
outer} halo or other  
 stellar associations. 
In fact several evidences suggest that the halo population
is contaminated by accreted stars (Majewski  1992; Preston, Beers
\& Shectman  1994; Beers \& Sommer-Larsen  1995; Carney \PBal 1996 ).
 Beers and Sommer-Larsen (1995)
 have shown that the disk population rotating at roughly 200 km/s 
 contains a  metal poor tail  extending down to [Fe/H] =-3 or  lower.
 They  also argued  that a significant fraction of metal poor stars 
 ($\approx$ 30 \%
 of the stars with [Fe/H]$<$-1.5) belongs to the thick disk.
  Such a population with metallicities typical of halo stars
 is probably   responsible for the 
 lack of a clear correlation between the kinematical properties
 and stellar metallicities in the Galaxy. 

 The nature of these very 
 metal poor stars with  disk rotational 
properties is not well understood.
 They may be   either the oldest stars  of the disk 
 or the result  of the collision of a preexisting thin disk
 with a dwarf satellite galaxy. These stars
  allow to establish 
 the chemical conditions at the onset of the
 stellar formation in the 
Galactic disk, in the former case, or in an old population
 of an accreted galaxy, in the latter case.
 Whichever the case they are a new important tool to verify the 
 pregalactic abundance of   lithium. 
 Molaro (1997)   already emphasized that
 the detection  of Li in a star belonging to the    Blue Metal Poor 
 population, discovered 
 by Preston, Beers \& Shectman (1994), possibly 
  resulting from  a
 merger event,  can be considered as an indirect  
 {\it extragalactic} Li  detection.

 In this paper we present lithium observations in seven
 main sequence 
stars spanning metallicities from -1.5  down to -3.0, 
but with
 kinematics belonging to the disk. 
 The observations 
 show that these stars have Li and that all, 
but one, share the same Li abundance with the genuine  halo dwarfs.
  
 \PBtbl{Journal of the observations
 }
 \halign to \hsize{\tabskip=0ptplus12ptminus15pt  
 # &
 # &
 # &
 # &
 # &
 # &
 #\cr
 \multispan{7}{\hrulefill}\cr 
 Star &  \hfill 
$\alpha$  & \hfill $\delta$  \hfill& \hfill V &  \hfill date  & \hfill 
 exp \hfill &  \hfill Instr \cr
 & \hfill 1950& \hfill 1950&\cr
  \multispan{7}{\hrulefill}\cr
 CS 29529-12 & \hfill 03 42 29.8  & \hfill  -60 29.7     & \hfill 12.0    &   \hfill  94 Oct 13     & \hfill2400+1800     & \hfill  E        \cr
 CS 29529-34 & \hfill 03 52 54.9  & \hfill  -59 15.2      & \hfill 13.9    &   \hfill  94 Oct 13   & \hfill 2$\times$3600       & \hfill  E        \cr
   & \hfill   & \hfill       & \hfill    &   \hfill  96 Sept 25     & \hfill 3600       & \hfill  C        \cr
 CS 22182-24& \hfill 04 18 03.9 & \hfill  -31 29.0     & \hfill 12.9   &   \hfill  94 Oct 13     & \hfill 3600       & \hfill  E        \cr
 CS 22186-50 & \hfill 04 30 21.1   & \hfill  -37 09.6     & \hfill 13.6    &   \hfill 94 Oct  13   & \hfill 2$\times$3600       & \hfill  E        \cr
  & \hfill    & \hfill       & \hfill    &   \hfill  96 Sept 25    & \hfill 2415      & \hfill  C       \cr
 CS 22191-17 & \hfill 04 34 49.9  & \hfill  -40 20.8    & \hfill 13.9    &   \hfill  96 Sept 25      & \hfill 2X3600       & \hfill  C       \cr
 CS 22959-7 & \hfill  18 36 02.1   & \hfill  -67 18.3     & \hfill 14.4    &   \hfill  96 Sept  24   & \hfill 2$\times$3600       & \hfill  C        \cr
 CS 22894-19 & \hfill 23 36 45.2  & \hfill  -00 12.9     & \hfill 13.9    &   \hfill  96 Sept 25     & \hfill 3$\times$3600       & \hfill  C      \cr
 \multispan{7}{\hrulefill}\cr
    }
 \endtable

 \beginfigure*{1}
        \centerline{
 \psfig{figure=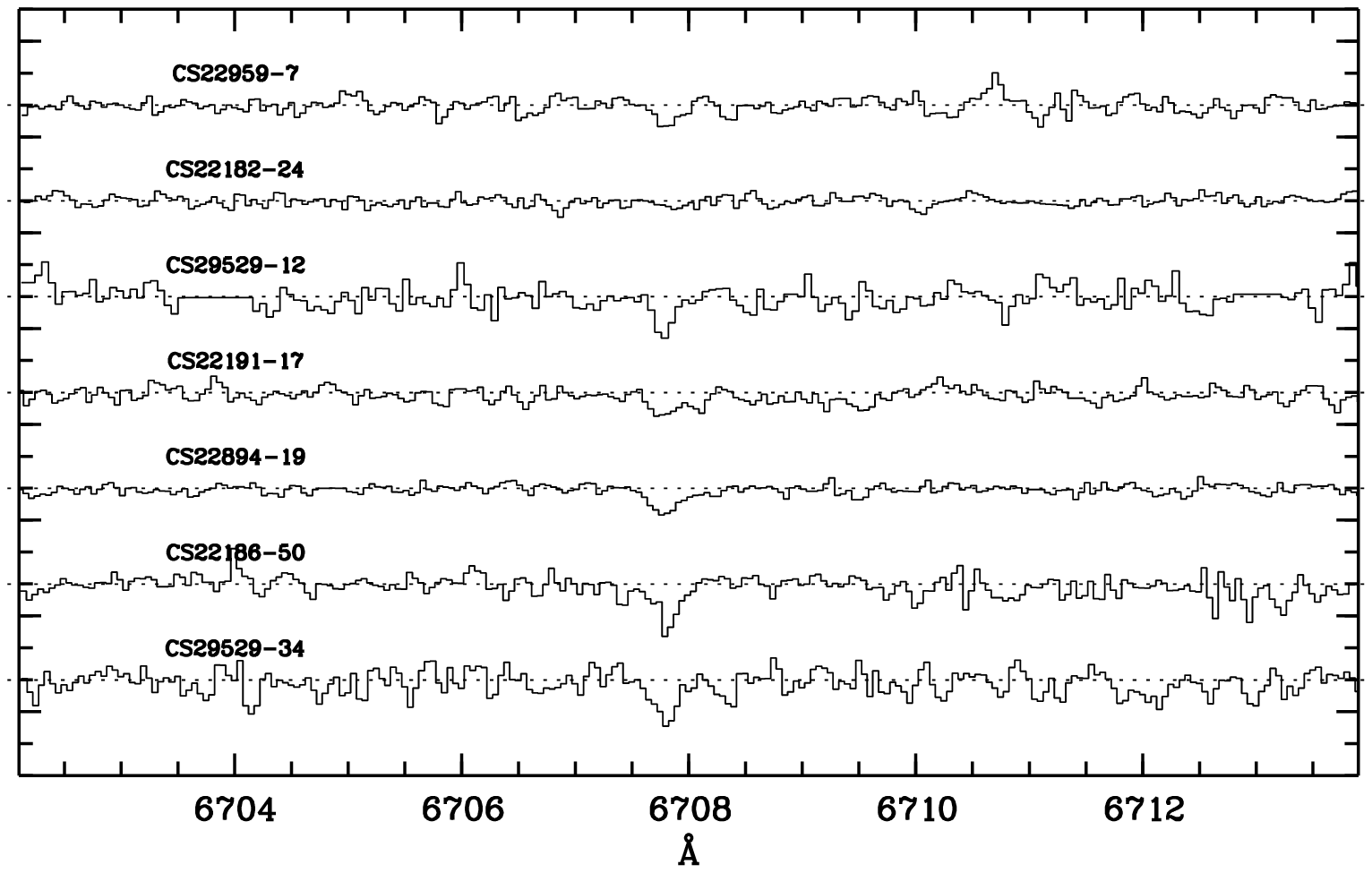}
         }
 \caption{{\bf Figure 1.} The Li region for the thick disk stars ordered
 from the top by decreasing (B-V). On the vertical axis
are plotted the residual intensities shifted by arbitrary amounts
for display purposes. Each tick on the vertical axis corresponds to
0.1 in residual intensity.  }
 \endfigure

 \section{Stars and observations}

 The targets have been selected among the most metal poor stars
 of the 
Beers and Sommer-Larson (1995) sample. The objects  are all turn-off
 stars with  [Fe/H]$\le$ -1.5, 
 and have disk-like rotational properties.
 To reduce the probability of  halo contamination the stars have been
 selected with  a distance on the 
 plane of $|z|\le$ 1 Kpc, 
 and in a position such that the line-of-sight 
 component of disk rotation is at least 100 km/sec.

 A first set of  observations was obtained
 between October 11 and October 14, 1994
 using the EMMI spectrograph  at 
 the Nasmyth focus of the 3.5m  NTT telescope. 
 EMMI  was 
 used in echelle mode, with grism 3 as cross disperser, 
 with  a Tektronix 2040$\times$2048 CCD detector and the  F/5 Long Camera
 the scale in the
 Li region
 was of   0.067 {\AA}/pixel. The    slit aperture was set at  $1.2\times
10$ 
  arcsec 
 providing  a resolving power of  
R= $\lambda / \delta\lambda$ $\sim$ 31000. 
 The slit height  gives 35 pixels
  perpendicular to the dispersion  direction  allowing  a satisfactory
 background-sky subtraction. 
   The CCD was  read in  
 super-slow mode with  a RON of $\approx$   4e$^-$/pixel. 
 A second observing run 
 was conducted on September  1996 using the CASPEC spectrograph
 at the Cassegrain focus of the 3.6 m ESO telescope. The echelle grating
 with 31.6 lines/mm and the long camera (f/3) were used.
 The detector was a Tektronix 
 CCD $1024\times 1024$ 
square pixels of 24 microns in size with a RON of 
  4e$^-$/pixel.
 With the  long camera the 
 scale at the detector was 
 of 0.077 {\AA} /pixel.  The slit 
was set at $1.2\times 10$ arc-seconds providing a
 resolving power of  
 R= $\lambda / \delta\lambda$ $\approx$ 32000 as measured from the
 Th-Ar emission lines.
  The journal of the observations is  summarized in Table 1.
  
  The data were reduced 
 using the {\tt MIDAS  ECHELLE}  package. 
 Special care was taken in the reduction in order 
 to  subtract  the background  contribution
 and to take into account the presence of high energy particles events. 
 The uncertainty in the wavelength calibration 
 is of $\approx$ 0.2 
of the resolution element,  or $\approx$ 2 km/sec. The 
 heliocentric radial velocities,  
 as measured from the NaI  lines are reported in Table 3. 
  The continuum was traced 
 measuring the intensity of the spectra in several windows  
  in selected  regions surrounding the Li line. For multiple observations 
 the spectra were  coadded with the proper weight after reduction
 to a common heliocentric scale.

 Equivalent widths were 
 measured from  gaussian  fit integration. 
 The errors on   equivalent widths
 have been quantified by taking $\sigma_{EW}$=K$\cdot$ (S/N)$^{-1}$,
 where K$\sim 1.503(F \cdot d)^{-1/2}$, where F is the spectral resolution
 and d the pixel size in \AA~ (Cayrel 1988, Deliyannis et al 1993). 
 With our setup K values are of 189 and 168 m\AA~ for
 Emmi and Caspec spectra, respectively.
 The 
  S/N ratio  is  measured in the continuum windows 
 close to the Li line.
  Equivalent widths with errors  are given in Table 3.  
 The final spectra are presented in Figure 1 after rebinning to equal 
 (0.05 {\AA}) wavelength step. 
  
 \PBtbl{Basic data for the stars.
 }
 \halign to \hsize{\tabskip=0ptplus12ptminus15pt  
 # &
 # &
 # &
 # &
 # &
 # &
 #\cr
 \multispan{6}{\hrulefill}\cr 
 Star &  [Fe/H] & (B-V) & \hfill E$(B-V)$ \hfill& \hfill E$(B-V)_{NaI}$  & \hfill 
 T$_{eff}$ \hfill   \cr
  \multispan{6}{\hrulefill}\cr
 CS 29529-12
  &  -1.54 & \hfill 0.40   & \hfill  0.00     & \hfill  $>$0.06  & 
        \hfill 6197                   \cr
 CS 29529-34
  &  -2.48 & \hfill 0.44   & \hfill  0.00     & \hfill  -  & 
        \hfill 5954                   \cr
 CS 22182-24
  &  -1.72 & \hfill 0.38  & \hfill  0.00     & \hfill  $>$0.02  & 
        \hfill 6259            \cr
 CS 22186-50 
  &  -2.02 & \hfill 0.43   & \hfill  0.00     & \hfill  -  & 
        \hfill 6014                 \cr
 CS 22191-17
  &  -2.13 & \hfill 0.39  & \hfill  0.00     & \hfill  $>$0.01  & 
        \hfill 6177          \cr
 CS 22959-7 
  &  -1.69 & \hfill 0.37   & \hfill  0.05    & \hfill  $>$0.06  & 
        \hfill 6541                \cr
 CS 22894-19
  &  -3.03 & \hfill 0.44   & \hfill  0.02     & \hfill  $>$0.05  & 
        \hfill 6060                   \cr
 \multispan{6}{\hrulefill}\cr
    }
 \endtable

 \PBtbl{Lithium abundances   }
 \halign to \hsize{\tabskip=0ptplus12ptminus15pt  
 # &
 # &
 # &
 # &
 #\cr
 \multispan{5}{\hrulefill}\cr 
 Star &   Vr$_{hel}$ & EW & A(Li) & A(Li)$_c$ \cr
  \multispan{5}{\hrulefill}\cr
 CS 29529-12 & 94  &\hfill 24 $\pm 6$ & \hfill 2.21 $\pm$ 0.17  & \hfill
2.24 \cr
 CS 29529-34 & 55 &\hfill 32 $\pm 7$ & \hfill 2.11 $\pm$ 0.15    &  \hfill 2.18 \cr
 CS 22182-24
  & 101 &\hfill 7 $\pm 3$  & \hfill 1.61 $\pm$ 0.40    & \hfill 1.63 \cr
 CS 22186-50  & 54 &\hfill 35  $\pm 5$ & \hfill  2.20 $\pm$  0.12  & \hfill 2.24  \cr
 CS 22191-17
  & 48 &\hfill 23  $\pm 4$ & \hfill  2.19 $\pm$ 0.13   & \hfill 2.21 \cr
 CS 22959-7 &59 & \hfill 17  $\pm 5$ & \hfill  2.21 $\pm$ 0.19   & \hfill    2.23 \cr
 CS  22894-19 &35 &\hfill 25  $\pm 3$ & \hfill  2.11 $\pm$ 0.11   & \hfill  2.17 \cr
 \multispan{5}{\hrulefill}\cr
    }
 \endtable

 \section{ L\lowercase{i} abundances}

 The stellar effective temperatures 
 have been derived from the (B-V)$_{0}$
 colour by using 
  the   calibration of  Alonso et al. (1996b). The
 temperature scale is  based on a set of 
 T$_{eff}$'s obtained by  Alonso et al. (1996a) 
 with the Infrared Flux Method (IRFM)
 and is  probably  the most accurate temperature scale for metal poor
 stars.
 It is 
 hotter 
 by about 100 K than previously used temperature scales.
 
  An independent evaluation of the 
 foreground reddening by inspection of the NaI interstellar 
 lines  has been performed 
 when the lines were not significantly affected by  NaI sky emission.
 The foreground reddenings  derived  by using the relations $\log $ 
 N(NaI)=1.04 
 [$\log$ N(H+2H$_2$)] -9.09 (Ferlet et al 1985) and 
 N(HI)/E$_{(B-V)}$=5.8 $\cdot $10$^{21}$ (Bohlin et al 1978)  
 are given in Table 2. The values  
 suggest that some of the stars might be in fact slightly more reddened
 than assumed and therefore  slightly hotter than  estimated here.
 A reddening of E$_{(B-V)}$=0.02 
 corresponds to  an increase of about 
 80 K in T$_{eff}$, i.e.   about 0.08 dex in A(Li).

  The programme stars are classified turn-off by
 Beers and Sommer-Larson 
(1995) and we have adopted  a gravity of $\log g$= 4.0. 
Surface gravity is not at all critical in the lithium
 abundance determination, 
provided the star has not evolved away from the 
 main sequence 
 and is therefore Li  diluted.  Microturbulence was fixed to 1.5 Km/sec
 as this value is common 
 for halo dwarfs. The precise value of microturbulence is also
 not critical for Li abundance determination.

 Li abundances are computed by using synthetic profiles computed by 
 the SYNTHE code and atmospheric models  with the ATLAS 9 code
 (Kurucz 1993). 
 ODF's computed with  $\alpha$ elements  abundances enhanced by  0.4 dex, 
 which better reproduce the chemical composition of Pop II stars,
 were used.
  Convection was 
 treated with the mixing length theory, with a mixing length of 
 1.25 the pressure scale height. 
 The   overshooting option, which   is
 implemented in  the convection treatment
 in the ATLAS 9 code, does not seem to produce better 
 agreement between computed and observed quantities 
 in stars other than 
the sun (Castelli Gratton \& Kurucz 1997) and it is not adopted here. 
 To quantify the difference:  if  the Kurucz (1993) 
 grid with overshooting were used, the 
 Li abundances 
 would be 
 increased by $\approx$ 0.05 dex. 
 The Li 
abundances are given in column 4 of Table 3. Corrections for 
NLTE according to Carlsson et al (1994) 
and expected depletions in the 
standard models following 
Deliyannis et al (1990) 
are taken into 
account in the Li abundances reported in column 5 of Table 3.

 \beginfigure*{2}
 \psfig{figure=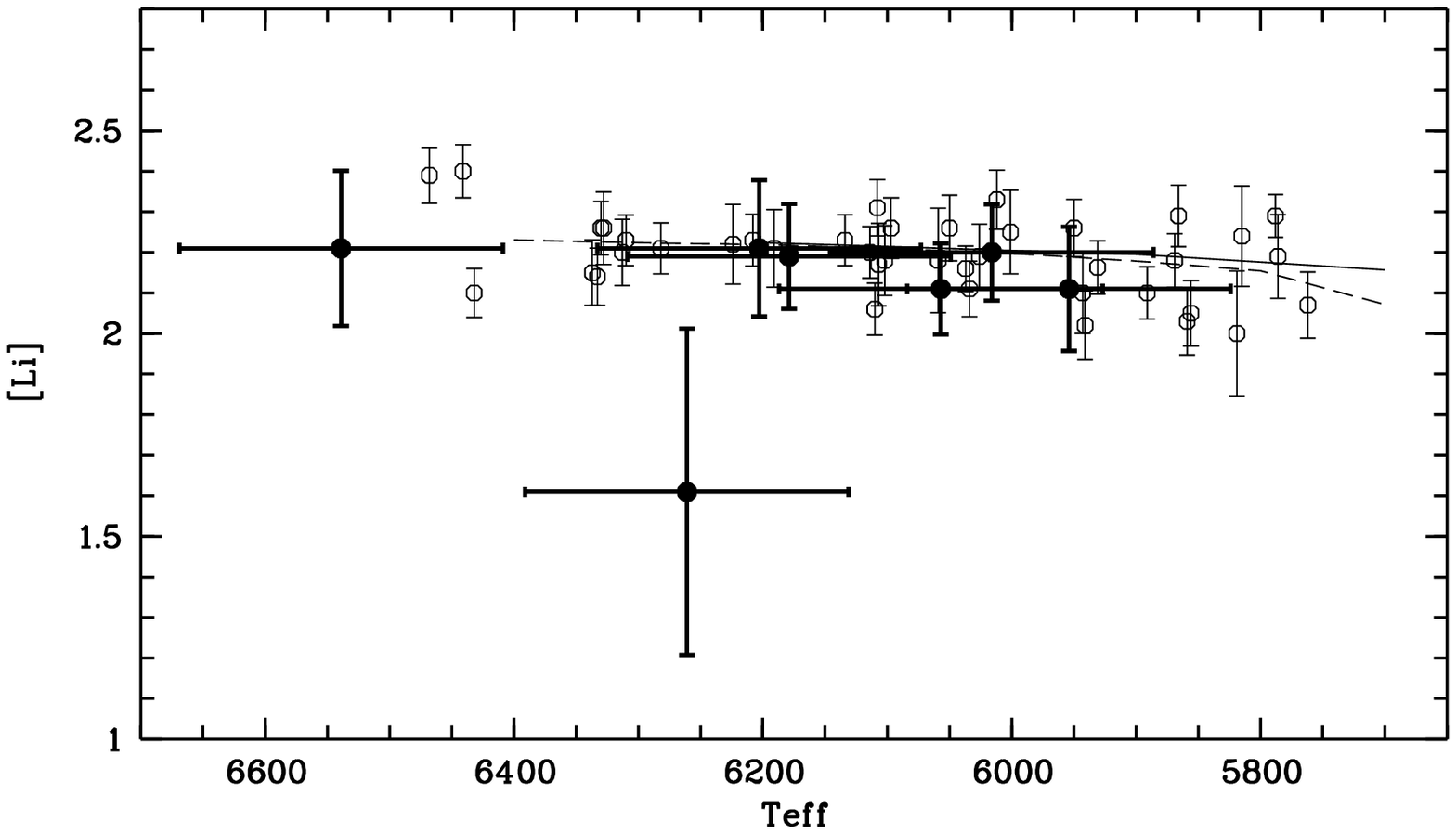}
 \caption{{\bf Figure 2. }Li 
abundances for the thick disk stars (filled circles)
 and field halo stars 
      from Bonifacio and Molaro (1997). Errors in T$_{eff}$
 for the comparison stars are omitted for clarity. The
 solid line is the ZAMS isochrone of Deliyannis \PBal (1990)
corresponding to [Fe/H]= $-1.3$ and an age of 16 Gyr, the dashed
line is the same but for [Fe/H]=$-2.3$. Both isochrones
have been shifted upwards by 0.1 dex to better match the observations.
  }
 \endfigure

 \section{discussion}

The formation of our Galaxy  is a process still to be understood in detail:
metal--poor stars can be either the relic of a rapid collapse or the
result of accretion of independent fragments (Carney \PBal 1996).
Analysis of rotational velocities and metallicities for metal poor stars
does not show a tight correlation as it is expected from classical 
rapid colapse models.
There is growing observational evidence that the old stars in the Galaxy
comprise several populations. Nissen\& Schuster (1997) found 
a group of halo stars ([Fe/H]$\approx$ -1.0) with solar values for the 
[$\alpha $/Fe] ratios. This group of stars is characterized by large
distances from the Galactic plane and  Nissen\& Schuster suggest 
they may be accreted from a dwarf galaxy with different chemical history.
Beers and Sommer-Larsen (1995) from an analysis of $\approx $  2000 
low metallicity 
 ([Fe/H] $\le$ -0.6) Galactic stars found that the galactic disk
contains a metal poor tail. As a part of a more general program of
Li  observations in the galactic populations we searched for Li in some of the
stars
 showing  disk-like rotational 
 properties but still having metallicities [Fe/H]$\le$ -1.5,
more typical of halo stars. In particular 
 CS 22894-19 has the   remarkably low metallicity of [Fe/H]=-3.03.
   
All these stars show a clear Li feature, with the  exception of CS 22182-24
where  the lithium line is detected  at a poor statistical significance
($\approx$ 95 \% C.L.)
The  Li abundances of the selected targets are shown in Fig 2 as a 
function of effective 
temperature. Also shown are the 
Li abundances for the 41 
halo stars having [Fe/H]$<$-1.5 and 
T$_{eff}$ $>$ 5700 K from Bonifacio and Molaro (1997). 
 The two sets of abundances 
 can be directly compared to each other because  they
 have been obtained by the same atmospheric codes and the stellar temperatures
are based on the 
 same IRFM temperature scale  of Alonso \PBal (1996a).
As it is clear  the Li abundances for the two different stellar 
populations largely overlap at all temperatures of the {\it plateau}, with the
exception of CS 22182-24.

It is likely that CS 22182-24  is truly Li-poor, 
 and is the  analogoue of the Li-poor stars found in the halo such as 
  G122-69, G139-8, G186-26 (Thorburn 1994) and 
G66-30 (Spite \PBal 1993). But other possibilities
cannot be ruled out. The star can
be more evolved of what assumed here 
or  be a binary. 
If the companion is of 
similar temperature the correction for the veiling  should double the EW, thus
rising the  Li abundance to the {\it plateau} level.

      The weighted mean of the 
Li abundances for all the seven stars corrected for NLTE effects and 
for the standard depletion  
is A(Li)$=2.20\pm 0.06$. The mean is only  increased  to  2.21 when
 CS 22182-24,  is removed as outlier.

This value  is slightly lower but fully consistent within less than 1 $\sigma$
with the {\it plateau} value of  A(Li)=$2.24\pm 0.012$
obtained by  
Bonifacio \& Molaro (1997) 
from the measurements of 
41 {\it plateau stars}, thus 
providing evidence that the metal poor stars with disk-like
 kinematics share the same Li abundance of 
genuine halo stars. 
An even  better consistency 
between the two values can be  
obtained when the 
stellar T$_{eff}$  are obtained from the dereddened colours assuming the
colour excess  derived 
from NaI interstellar lines, which result in somewhat larger Li abundances.

 It is not  clear whether  these stars  belong to the  first stellar
  disk generations 
 or whether they are accreted by the Galaxy. Thus they
  either probe
 the chemical conditions at the onset of the
 stellar formation in the Galactic disk or
     in an old population
 of an accreted dwarf galaxy. 
Whichever the case they allow to probe  an environment
which is different from the Galactic Halo.
The  fact that the two distinct metal-poor populations share the same Li
abundance
  provides a new  strong evidence 
that the pregalactic Li abundance was uniform and close to the 
value presently observed in the Galactic halo stars.

\section*{Acknowledgments}

 We wish to thank  T.C. Beers for checking out the  colors
of star CS 22182-24.

 \bye